\documentclass[fleqn,10pt]{wlscirep}
\usepackage{epsfig}
\usepackage{times,mathptmx}
\usepackage{amsmath} 
\def \be {\begin{equation}}
\def \ee {\end{equation}}
\title{Arrays of coupled chemical oscillators}

\author[1,*]{Derek Michael Forrester}
\affil[1]{Loughborough University, Chemical Engineering, Loughborough, LE11 3TU, United Kingdom}

\affil[*]{d.m.forrester@lboro.ac.uk}

\begin{abstract}

Oscillating chemical reactions result from complex periodic changes in the concentration of the reactants. In spatially ordered ensembles of candle flame oscillators the fluctuations in the ratio of oxygen atoms with respect to that of carbon, hydrogen and nitrogen produces an oscillation in the visible part of the flame related to the energy released per unit mass of oxygen. Thus, the products of the reaction vary in concentration as a function of time, giving rise to an oscillation in the amount of soot and radiative emission.  Synchronisation of interacting dynamical sub-systems occurs as arrays of flames that act as master and slave oscillators, with groups of candles numbering greater than two, creating a synchronised motion in three-dimensions. In a ring of candles the visible parts of each flame move together, up and down and back and forth, in a manner that appears like a ``worship''. Here this effect is shown for rings of flames which collectively empower a central flame to pulse to greater heights. In contrast, situations where the central flames are suppressed are also found. The phenomena leads to in-phase synchronised states emerging between periods of anti-phase synchronisation for arrays with different columnar sizes of candle and positioning. 

\end{abstract}
\begin{document}

\flushbottom
\maketitle
\thispagestyle{empty}

%\noindent Please note: Abbreviations should be introduced at the first mention in the main text – no abbreviations lists. Suggested structure of main text (not enforced) is provided below.

\section*{Introduction}
For $one-million$ years hominins have had the ability to control fire\cite{Berna2012}, but still there is much to learn about it. During the last century people have carefully controlled the spread of fire in the forests around the world. Fires propagating through natural processes, especially those generated by lightning strikes, would have in the past been limited to the destruction of small trees and bushes. However, a policy of suppressing almost all fires has resulted in the new phenomenon of the ``mega-fire''\cite{megafire2013}. The mega-fires are sustained by a build up of fuel, composed of small trees and shrubs, that provide a ladder for the fire to move into the forest canopy. This leads to intense, almost uncontrollable crown fire. Techniques have emerged to combat the spread of fire, including the use of fire cans to light controlled fires to take away the shorter forestation, but rising  costs of fire suppression calls for new methodologies.  

In order to better understand the dynamics of fire, particularly the proximity effects of individual flames that arise and are perpetuated by a fuel source, arrays of candles are examined for which phase locking \cite{BalanovJensen} of the flames is a feature. These flames act as $N$ coupled oscillators. The flames of the candles are found to synchronise, show small periods of phase drift, and have an interspersion of oscillation death (and hence order) where the oscillations cease. A candle flame has a perhaps surprising complexity, with still many details unknown to date about the combustion chemistry and internal composition. The pathway of a complex chemical reaction has many changes in the concentration of its components as it proceeds. 

Chemical oscillations have been found in a number of different systems. The oscillating Belousov-Zhabotinsky chemical reaction in-particular has been investigated extensively\cite{DolnikEpstein1996}.  The chemical oscillators composed of  arrays of candle flames investigated here,  are open systems in which the reactants and products are transported as part of a molecular diffusion process. There are various reaction zones within the flames in which carbon, oxygen and hydrogen species diffuse and react. The external reservoir outside the flames is the air of the environment and it replenishes the supply of oxygen which diffuses into the reaction zones continuously to produce intermediaries and products. The emission of light is synonymous with a flame and it is from observing its nature that many important insights into the fundamental characteristics of matter have been made. A diffusion flame has a constantly changing chemical composition from top to bottom and the burn rate is a function of the rate of diffusion and mixing\cite{Gaydon}. Single diffusion flames have been known to flicker at $10-12Hz$\cite{Grant}, with explanations due to buoyancy \cite{Buckmaster1986} with the region close to the base of the flame working as a source of perturbation to the whole plume\cite{Maxworthy}. Large-eddy simulations can be used to offer insight into the dynamics of the flickering of flames and have been used in the past to elucidate vortex dynamics and show that self-sustained flickering has a dependency upon shear-layer thickness\cite{vortex}. Toroidal vortices driven by buoyancy are also known to stretch the flame and create a detached puff \cite{Goss1989}. Thus, we extend the investigation of candle flicker to a demonstration of the synchronous behaviour evident in the chemical oscillations observable through the flames of arrays of candles.   
\begin{figure}[ht!]
\centering
\includegraphics[width=\linewidth]{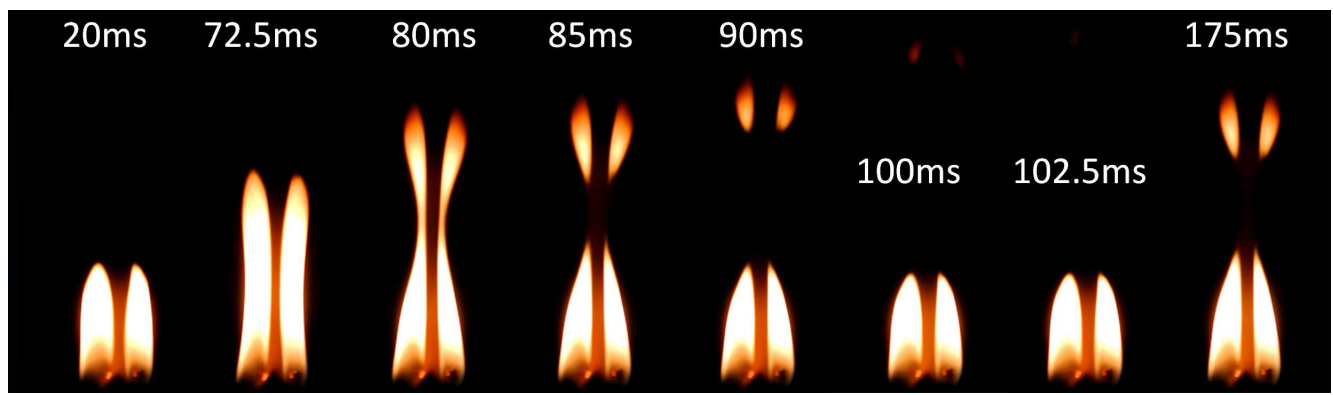}
\caption{The visible luminosity of the flames of two paraffin candles in synchronisation. Footage can be found in the Supplementary Information, ``2 candles.mov'' 
\label{fgr:Fig2Candles}}
\end{figure}
 
\section*{Results}
Figure \ref{fgr:Fig2Candles} shows a number of the properties associated with the synchronised states, demonstrated with two candles. The chemical oscillations \cite{Epstein2014,Takayama2013} in the combustion of the two candles are at a stage in Fig.\ref{fgr:Fig2Candles} where they are phase locked. The separation distance between the centres of the paraffin candles was $2cm$ and the diameters of the candles was also $2cm$. In all experiments a Nikon J1 camera captured the dynamics at $400fps$ so that the total number of frames (for recordings of $5s$ duration) was $2000$. The playback from the footage runs at $29.97fps$. The footage of the oscillations can be found in the Supporting Information. For two candles one can see over $20-175ms$  that the candle flames move together, gaining amplitude till a critical level where the visible part of the flames separates into two parts. Work on two coupled clusters each consisting of three candles has demonstrated similar phenomena \cite{Kitahata2009}, though the puffing of the flames was not focused upon. One can see a ``necking effect'' to the flames at $80ms$ and the splitting of the flames into two components at $85ms$. Two smaller visible flames continue the upwards trajectory until disappearing from the visible spectrum, whereas the main components have begun to move back down towards the candles. We will see that this splitting introduces anti-phase oscillations into the dynamics of larger arrays. The chemical composition of the flames themselves is highly complicated. Recent work discovered that in a candle flame millions of nano-diamonds are produced, along with fullerenes, graphite and amorphous carbon \cite{Soot2011}. Forest wildfires are also reported to be a source of nano-diamond \cite{Nanodiamonds2012}. The details of the chemical reactions can only be understood through quantum mechanics\cite{Atkins1991}. It is from the excited electronic states of the $OH$, $CC$, and $CH$ radicals that emission of discrete radiation bands occurs, giving the visible luminosity. 
\begin{figure}[ht!]
\centering
\includegraphics[width=\linewidth]{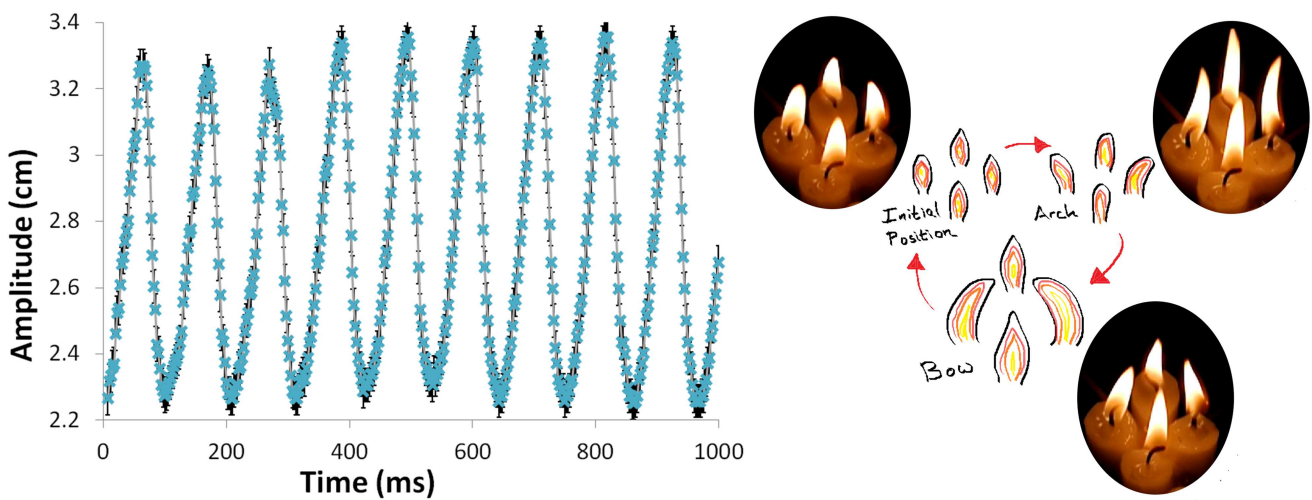}
\caption{A ring of four candles, with centre-to-centre spacing of $2cm$, presents synchronised oscillation of the visible luminosity. Left: The oscillation of the left-most candle flame is outlined as a function of amplitude (rise of the visible flame from the position where the wick connects with the pool of wax to its maximum height in space). Right: Three major stages emerge - $(1)$ the initial position, $(2)$ arching backwards and $(3)$ a dramatic bow (shown schematically with a snapshot from the footage alongside). Footage can be found in the Supplementary Information: ``4 candles.mov''.  \label{fgr:Fig4Candles}}
\end{figure}
The intensity of the yellow and red spectral elements is higher than the blue, and there is a distinctly visible oscillation of the continuous emission at these spectral wavelengths. This implies that there is a continuous extra flux of the carbonaceous particles, of size around $10-100nm$, as observed through the movement of the characteristic yellow luminosity. Each particle acts as a nano-sized grey-body\cite{emissivity1960}.    
\begin{figure}[ht!]
\centering
  \includegraphics[width=\linewidth]{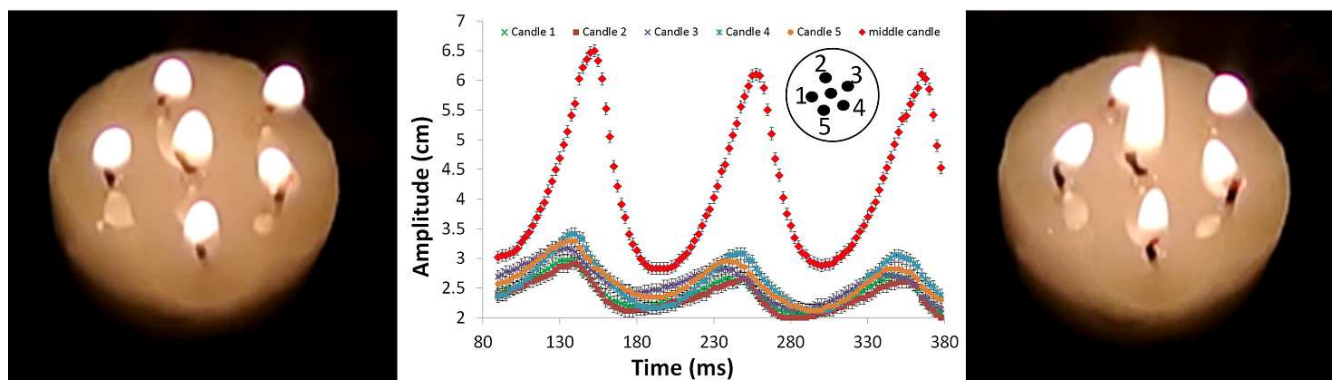}
  \caption{A pentagonal outer ring of candle flames enslave a central flame, enhancing its amplitude and perpetuating oscillation. Video of the effect is in the ESI as ``6 candles.mov''.  \label{fgr:Fig6Candles}}
\end{figure}  

In the Supplementary Information footage of three candles separated by $2cm$ in a triangular pattern can be found as they go between synchronised oscillation and oscillation death. Even the simplest example of two oscillating candle flames can produce a rich physical chemistry, with the oscillation described by non-linear coupled equations with a dependency on thermal radiation, separation distance, the fuel source, oxygen supply, and to a lesser extent convection currents\cite{Kitahata2009}. Indeed in the Mach-Zehnder interferometry experiments of Ref. \cite{Kitahata2009} convection was determined only to have a cooling effect without causing phase locking. However, given the intermittent occurrences of the synchronised motion of the visible luminosity in ``3 candles.mov'', one cannot dismiss convection and vorticity effects as the reason of termination of the oscillation, though local oxygen fluctuations seem likely to be inter-related with the source of restoration. It is also possible that the separation distance is at a critical level because moving the candles closer increases the regularity of the synchronisation.

To investigate group effects due to coupling, larger clusters and arrays are now shown. Four candles display a very interesting characteristic of synchronised motion that is now described through Fig. \ref{fgr:Fig4Candles}. Starting at a low amplitude, the four candle flames synchronise and stretch out. This can be seen in the amplitude versus time plot in the top of Fig. \ref{fgr:Fig4Candles}, obtained from digitisation of the frames from the slow-motion footage. During these oscillations the visible parts of the flames appear to rise and arch backwards before bowing together. This characteristic is not seen in the coupling of two candles but was also in evidence to a lesser degree for three. The behaviour can be seen in ``4 candles.mov''. As a network the four candles are arranged within a closed loop, which serves to perpetuate the arching/bowing phenomenon. In Fig. \ref{fgr:Fig4Candles}, as by way of example, the time from the initial to the arch position is close to $35ms$ (e.g. time $210ms$ to $245ms$). The time is approximately the same for the two other stages (arch to bow: $245-280ms$; bow to initial: $280-315ms$), with the whole initial-arch-bow-initial ``worship'' cycle taking $105ms\pm 3$. In contrast two candles of the same type, size and separation distance (as in Fig.\ref{fgr:Fig2Candles}), will complete an oscillation cycle in $87ms\pm 3$. For completeness, five candles are found to have the same ``worshipping'' appearance when arranged in a circle (``5 candles.mov'') but also with a break of symmetry to have one adjacent candle next to a cluster of four (``5 candles symmetry break.mov''), in the Supporting Information.

Next it is instructive to examine a ring of candles with one candle positioned in the centre. A pentagonal arrangement is made around the central candle in Fig.\ref{fgr:Fig6Candles} and the characteristic synchronisation is in evidence again (``6 candles.mov''). However, the five outer candles now accentuate the effect on the central candle, enhancing the height to which its flame is pumped. The flames of each candle oscillate almost synchronously with small delays between each maximum amplitude, as can be seen in Fig. \ref{fgr:Fig6Candles}, which are not noticeable until examining the slow-motion footage in detail. The motion of the oscillation is akin to producing a braiding effect in one candles flame profile that weakly twists over and under the amplitude signature of its neighbours as a function of time (see plot in Fig.\ref{fgr:Fig6Candles}, for candles $1-5$). With small perturbations to the system, it is required that it readjusts itself, meaning that it is a self-sustained effect that is robust against small disturbances. The system is homogeneous enough for synchronisation to prevail.

\begin{figure}[ht!]
\centering
  \includegraphics[width=\linewidth]{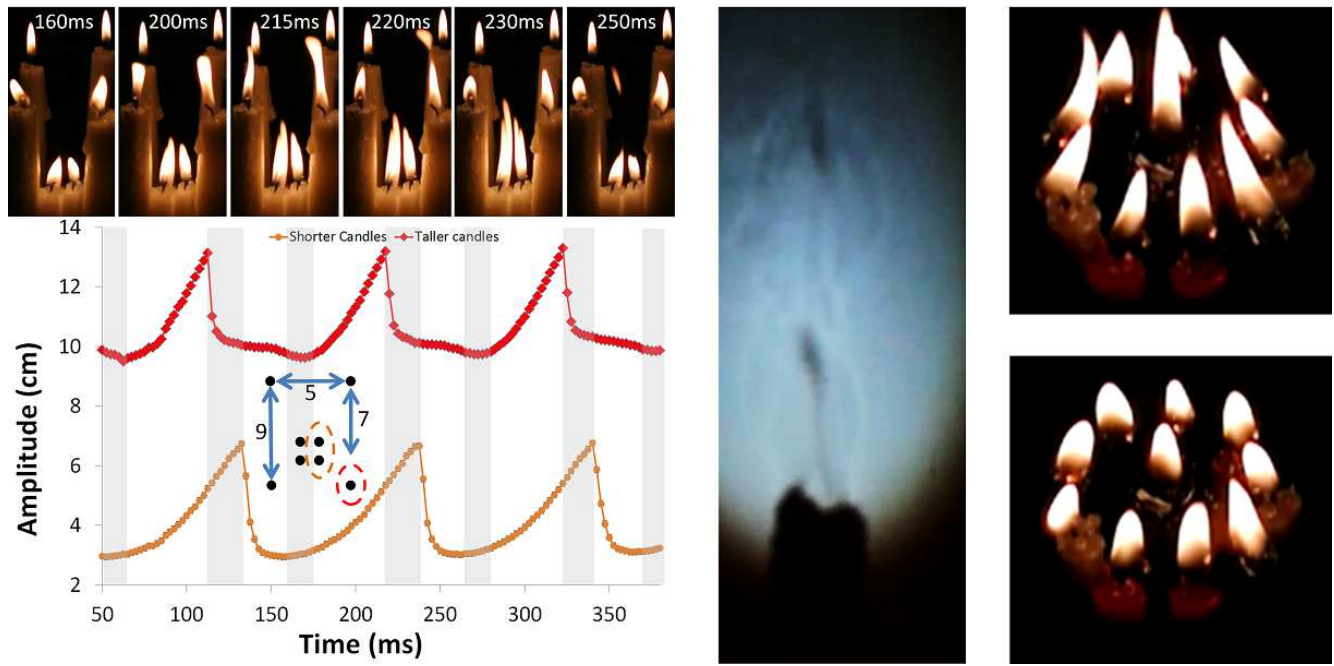}
  \caption{Left:Eight candles in an array, each of $2cm$ diameter, where a central cluster of four flames drives the oscillation, whilst coupled to two nearest-neighbours. The two candles furthest from the cluster are stable. The largest amplitude of the two right-hand visible flames of the cluster is plotted along with that of the taller right-hand outer candle closest to the cluster. Snapshots from the footage in the ESI (``8 candles.mov'') are shown above. Centre: Shadowgraphs demonstrate the wave of particles as dark regions in the centre line of the flame (footage in the supporting information). The ascending oscillating of the hydrocarbon vapour from the point of the wick is evident. Right: Ten coupled oscillators. \label{fgr:FigHeight}}
\end{figure}

Figure \ref{fgr:Fig6Candles} shows that the pentagonal distribution of candles serves to approximately double the height of the central visible luminosity. In the Supplementary Information further six candle arrangements with a ladder and line layout can be found ``6 candles ladder.mov'' and ``6 candles line.mov''.  In the ladder arrangement the two flames of the middle ``rung'' have a greater amplitude. It is rather interesting that when the experiments are repeated for a hexagonal arrangement of candles surrounding a central one, that the natural condition was found to be that the outer candle flames pulse higher, and suppress the amplitude of the middle flame. The order in which the candles are lit also makes a temporary difference and causes an adjustment of the balance to the system. When the central candle was extinguished and then relit a competition arose between it and two side-by-side outer flames, forming a triangle. Further, if one of the remaining four outer candles began to compete too (usually next-nearest-neighbours to the two original competitors on the outside), the overall amplitude of each candle equilibrated to a large degree, with the central one usually lowest (``7 candles.mov''). With nine outer candles the effect is almost identical to the six candles with a pentagonal outer candle arrangement (see Fig.\ref{fgr:FigHeight} and ``10 candles.mov'').
	
Arrangements of candles where a cluster of four are surrounded by a group of taller candles show that the synchronisation can be extended to happen over larger distances by use of the cluster as an intermediary. Shadowgraphs (``shadows.mov'' series) for a cluster of four candles are shown in the Supplementary Information, performed as per Faraday \cite{Faraday} (who lectured about candle phenomena in the nineteenth century and used this technique). One can gleam information on the movement of the carbonaceous particles as the dark regions in the shadow as flicks of darker zones.  One can also see that above where the visible part of the flame is found, that there is a swirling mass leading to turbulent flow. Using these groups of four candles with their flames in synchronisation,  and placing them between taller candles, gives an effect whereby the surrounding group moves into and out of in-phase synchronisation. This can be seen in Fig.\ref{fgr:FigHeight} and ``8 candles.mov''. This behaviour occurs when the position of the cluster is made closer to one pair of outer candles. The furthest pair of candles from the cluster then attain stability. In Fig.\ref{fgr:FigHeight}, one can see that as the cluster and closest two neighbours oscillate, there is a critical point at which the visible luminosity splits into its two components, and the system enters into periods of anti-phase synchronisation. The amplitudes of the visible luminosity are no longer increasing together, but the closer outer candle flames amplitudes take an opposite gradient to the inner cluster. This is marked by a massive decrease in amplitude in Fig.\ref{fgr:FigHeight}, as it is measured from the height of the flames extended directly from the position of the lowest part of the visible wick of the front, right candle of the cluster. The maximum amplitudes of the two right-hand-side flames of the cluster are plotted against the amplitude of the right-hand-outer-front flame. As the flames evolve they stretch upwards continuously before splitting, with the main component beginning to vector downwards. In the ESI one can find in ``8 candles 2.mov'' that when the geometrical position of the array becomes symmetric, i.e. the cluster of four candles moves central, that the outer candles fully synchronise even separated by $9.5cm$. This is achieved with the influence of the cluster, and implies that extended group effects arise depending upon positioning. Further, in ``10 candles 2.mov'' the cluster of four lies between the central candles of a ladder arrangement of six. The outer flames, of candles separated by $9cm$ are stable, whereas the central outer candles have a longer period of anti-phase synchronisation than in the eight candle example. Thus, there are many delicate balances to the interaction of the candles that needs exploration.

\section*{Discussion}
Here, it has been demonstrated that a flame with a constant fuel source will form complex chemical coupling with other oscillators. The candles used to demonstrate the phenomena can be of different heights, and symmetrical or random positioning. Within a natural fire it is not unreasonable to expect that similar effects may occur, and for master/slave oscillator systems to at times control the spread. Indeed, forward pulsations of propagating wildfires have been commonly seen\cite{forwardpulsation}.
One can foresee that these effects could be used to entrain natural fires by careful design of new fire defences. For example, it may be possible to arrest the progression of the fire by preparing lines of periodically positioned lit ``candle-like'' poles. This is of course a large experiment that should be conducted by fire services in a controlled area. Knowing that arrays of fire can take over the dynamics of the global system brings into light a potential unexplored direction for fire fighting, perhaps delaying the progression of a fire long enough for effective measures to be taken to extinguish it.  In this work we have outlined the emergence of synchronisation events for larger arrays of candles than previously examined, finding new group effects. In addition, on-going work has indicated that larger arrays with a novel ``lattice'' structure exhibit periods of chaos, whereby the candles flicker randomly and out of sync. Inside the larger arrays synchronised ``domains'' can emerge whilst other neighbours remain asynchronous. It is possible that hybrids of synchronised and incoherent oscillators in the arrays - chimera states\cite{Nkomo2013} - could be witnessed. 

In related phenomena, under controlled circumstances plane laminar flames have been experimentally observed to become unstable and divide into cellular structures known as cellular flames (see for example Ref.\cite{Sivashinsky}). Markstein demonstrated the dynamics of cellular flames\cite{Markstein,MarksteinSomers} and Gorman \textit{et al.} found ordered patterns in heavy hydrocarbon-air mixtures\cite{Gorman1}. Although cellular flames are not observed in our experiments, it may be that premixed flames that have controllable flow rates of reactants could provide additional insights for understanding the complex dynamics. Indeed, diffusional-thermal flame \cite{Sivashinsky} reasons for cell formation and perturbative hydrodynamical effects (Darrieus–Landau instability\cite{Landau1944, Matalon})  have been previously stated as giving rise to different kinds of oscillatory behaviour\cite{Class,Sivashinsky}. The formation of rings of cells inside a flame is a function of the flow rate and the ratio of the fuel to oxidiser in these alternative kinds of premixed state experiments\cite{Gorman1}. In the experiments of Gorman \textit{et al.} periodic pulsating flames were observed in the cellular radial modes (and axial and drumhead modes), with the flame front expanding and contracting to alter the radial extent\cite{Gorman2}. Transitions to chaos also occurred as departures from an extinction limit were made\cite{Gorman2,Gorman6} (temperatures at crests of cellular flames are relatively low with a tendency towards extinction\cite{Sivashinsky}). Chaotic regimes of cellular flame dynamics can be described using Kuramoto-Sivashinsky modelling\cite{Class,Blomgren} (and many other reaction-diffusion effects as well as long waves on thin films\cite{Lakestania}, etc) and may also describe candle flame oscillations and synchronisation. 

The cellular flames appear as regions of bright valleys interspersed between darker peaks\cite{Gorman3}. In order to examine these cellular flames Gorman \textit{et al.} used a laminar premixed flame on a custom built circular porous plug burner, observing order\cite{Gorman1}, disorder\cite{Gorman2,Gorman4}, hopping motion\cite{Gorman5,Gorman7}, pulsation\cite{Gorman4}, and ratcheting\cite{Gorman8,Gorman9,Gorman10}. From the side they observed a solid sheet of flame with the boundaries of the "cells” formed by cusp shaped sinks and valleys, but from above a pattern of bright internal domain structures was clear\cite{Gorman1} (see Ref.\cite{Kadowaki} for a $3D$ depiction). These domains moved internally as a function of flow rate, with individual cells changing position and the number of cells fluctuating accordingly in a geometrical pattern. Observed from the top these cellular flames have similar geometrical patterns to the circular geometries of individual diffusive flames examined here at each candle. However, the phenomena examined here are quite different in origin: diffusive rather than premixed flames; with a positional dependency of the candles that is not evident in the cellular flames; perhaps a more generic flame oscillation (the cellular effects have only been observed for certain mixtures, e.g. isobutane-air flames\cite{Gorman3}); and a tendency for in-phase oscillation rather than hopping (i.e. in the cellular flames, inner cells will rotate in position as they compete for stability, whereas the candle flames are completely localised and do not propagate - for fire control this is essential as one would not want to add momentum to the spread of wild-fire). A momentary deformation of a cells shape moves its position relative to other cells that remain fixed in place. The ``worship'' cycle of the candle flames always happens with the flames moving in unison, not independently, even with small perturbations to individual flames. One might be able to use circular porous plug burners to stabilise flames\cite{Class,Buckmaster}, such as those used in the cellular flame experiments, to find similar effects to the candle experiments as a function of pore sizes and separation distances. Interestingly, Gorman \textit{et al.} found that when twelve outer cells appeared a centrally formed cell spiralled\cite{Gorman3}. When eighteen outer candle flames were circularly positioned in lines of three outside a central candle, the flame of the central candle spiralled as the six groups bowed slightly out of phase. Thus, it is possible that analogous behaviour between these systems does exist and could be complementary for finding further novel phenomena.

Another avenue of this research is the potential for collection of nanoparticles\cite{Soot2014}. With it well known that there are formations of nano-diamonds and other carbon allotropes in the flame, methods for their capture can be further designed. The oscillation of the visible luminosity associated with the nanoparticles presents the opportunity to create thin films on substrates entered into the flames, giving controlled thickness (calculations of the period of oscillation suggests the prospect for integer numbers of depositions as a function of collective ``worships'') in a manner similar to chemical vapour deposition. Thus, the grouped effects of these systems can perhaps lead to new techniques in the laboratory or greater understanding of fire dynamics for fire fighting in this age of the mega-fire. It may be possible that lines of control elements, analogous to candles, could be used to alter the trajectories of fire and guide it away from populated areas. Oxygen depletion in the vicinity of the oscillators could serve to align a spreading fire along an alternative pathway, providing a last resort barrier defence.  

\section*{Methods}

The relative distance between the tip of the visible flame and the wax pool surface was examined using the frame analysis features of Media Player Classic; a free software that enables frame capture. The dynamics of the flames were recorded using a Nikon J1 camera in the slow motion mode, enabling $400fps$ to be captured. Further programming then digitised the images and recorded the relative heights of the visible parts of the flames. Paraffin candles, as well as those made from bees wax, were used - both demonstrating the phenomena, ruling out possible impurity effects. Shadow graphs were recorded in slow motion as a beam of light was shone through the oscillating flames against a white background in a darkened room.

\section*{Acknowledgements}
This work has been supported by the EPSRC KTA grant - ``Developing prototypes and a commercial strategy for nanoblade technology''. The data reported here can be obtained by contacting the author. The author thanks Professor Kay Robbins for helpful correspondence with regards to cellular flames and the reviewer for highlighting the work of Professor Michael Gorman and co-workers.   
 
%\section*{Author contributions statement}

\section*{Additional information}

\textbf{Competing financial interests}. The author declares no competing financial interests. 

\end{document}